\documentclass{article}
\usepackage{spconf,amsmath,graphicx}
\usepackage{textcomp}
\usepackage{etoolbox}
\usepackage{booktabs}
\usepackage{url}
\usepackage{tabularx}
\usepackage[table]{xcolor}

\title{BC-VAD: A Robust Bone Conduction Voice Activity Detection}
\name{
    Niccolò Polvani$^{1,2}$*\thanks{* Niccolò Polvani performed this work as an intern at Logitech.},
    Damien Ronssin$^{2}$,
    Milos Cernak$^{2}$
}
\address{
    $^{1}$\'Ecole Polytechnique F\'ed\'erale de Lausanne (EPFL), Lausanne, Switzerland,\\
    $^{2}$Logitech Europe, Lausanne, Switzerland
}
\begin{document}
%
\maketitle
\begin{abstract}

Voice Activity Detection (VAD) is a fundamental module in many audio applications. Recent state-of-the-art VAD systems are often based on neural networks, but they require a computational budget that usually exceeds the capabilities of a small battery-operated device when preserving the performance of larger models. In this work, we rely on the input from a bone conduction microphone (BCM) to design an efficient VAD (BC-VAD) robust against residual non-stationary noises originating from the environment or speakers not wearing the BCM. We first show that a larger VAD system (58k parameters) achieves state-of-the-art results on a publicly available benchmark but fails when running on bone conduction signals. We then compare its variant BC-VAD (5k parameters and trained on BC data) with a baseline especially designed for a BCM and show that the proposed method achieves better performances under various metrics while keeping the real-time processing requirement for a microcontroller.

\end{abstract}
\section{Introduction}
\label{sec:intro}


Voice Activity Detection (VAD) is a module that detects speech in audio signals. It is fundamental in many speech-related tasks, such as speech enhancement, automatic speech recognition, emotion classification, etc. Recent improvements in voice activity detectors were proposed for novel platforms, like analog~\cite{Minhao18} or neuromorphic~\cite{Martinelli2020, Dellaferrera20}, while some other works proposed personalized VAD~\cite{Wang22icassp}. However, all those methods considered traditional analog or digital air microphones.

Further improvements can be achieved with novel MEMS bone-conduction microphones (BCM), which are inherently more robust against external noise than traditional air-conduction microphones (ACM). Due to its characteristics, auxiliary signal from a BCM has been used to improve the performance of speech enhancement systems \cite{BC_Noise_Estimation, DSP-VAD, BC_Speech_Enhancement_2020}, as well as pitch determination \cite{Pitch_determination_BC}, apriori SNR estimation \cite{A_priori_SNR_Estimation_BC}, human sound classification \cite{human_sound_classification} and so on. The BCM suppresses not only external noise but also interference speech from individuals that are not wearing the microphone. The BCM signal is, therefore, a suitable and sufficient input to perform personalized VAD. In particular, it does not require an enrollment phase from the user, unlike most personalized VAD \cite{personal_vad} or personalized speech enhancement systems \cite{Personalized_PercepNet}. 

As BCMs require close contact with the user's body, they are usually placed on small devices such as earbuds or headsets with limited computational resources. State-of-the-art air microphone VAD models dealing with non-stationary noises are large ($\approx$100k parameters) and would not fit on such small devices. On the other hand, existing BCM-based VAD systems do not consider leakage of external noise in the bone conduction sensor; they either ignore noise~\cite{BC_Noise_Estimation} or consider only thermal noise~\cite{DSP-VAD}.



In this work, we designed a tiny bone conduction VAD system (BC-VAD) that considers the presence of external noise by leveraging techniques used in air microphone-based systems. We first developed an ACM-based non-personalized VAD (AIR-VAD), achieving state-of-the-art performance on a public VAD benchmark~\cite{QUT-NOISE-TIMIT}. Then, we trained a smaller version of the previous model on our collected bone conduction data. We show that the developed method is robust against external noises and interference speech and performs better than an existing voice activity detector for BC. We finally evaluated the memory footprint and latency of BC-VAD by deploying it on an ARM-Cortex M33.

\section{Methods}
\subsection{Proposed model architecture}


\begin{figure}[htb]
  \centering
  \centerline{\includegraphics[width=8cm]{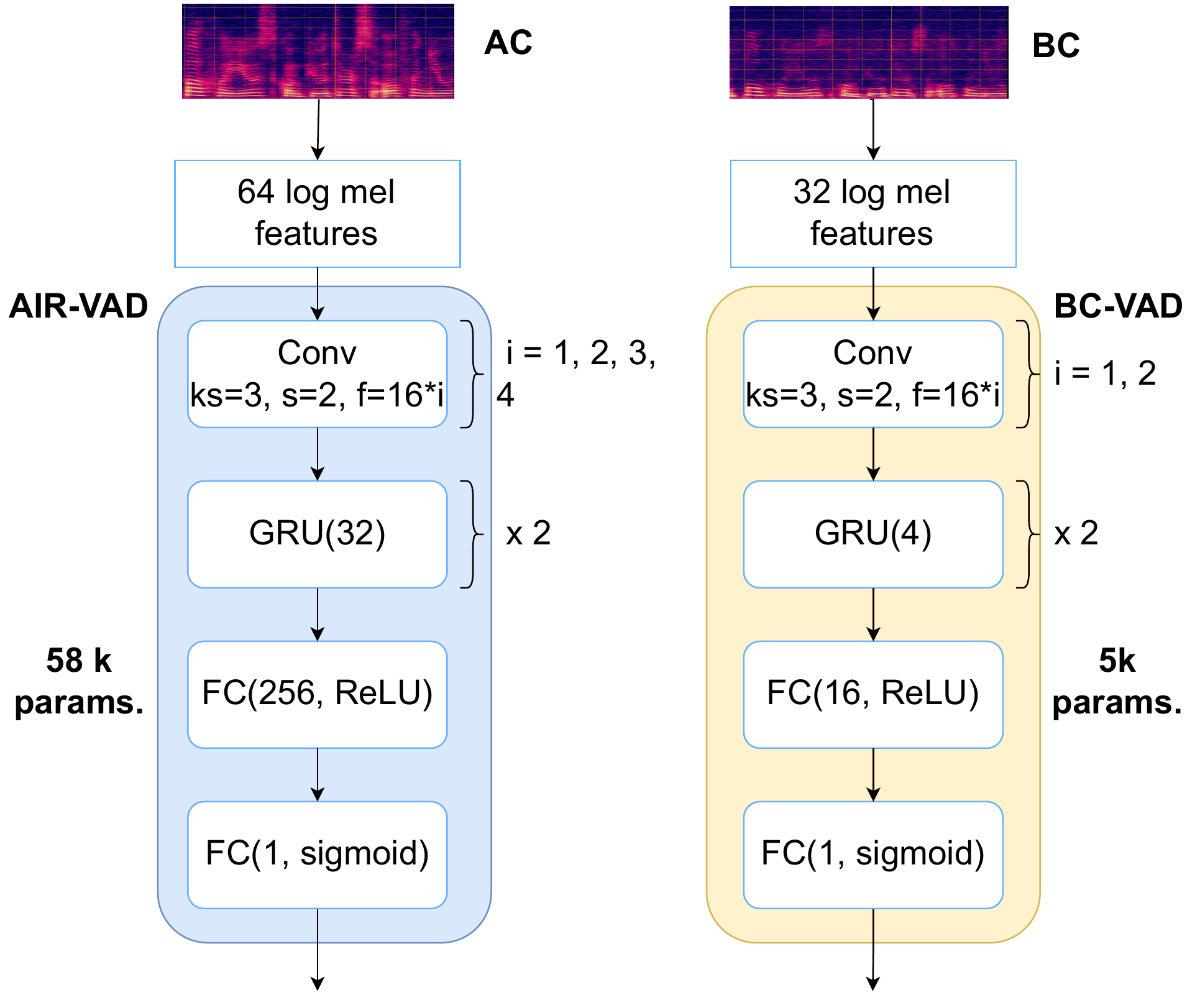}}
  \caption{Model architecture for AIR-VAD (left) and BC-VAD (right). $ks$ stands for kernel size, $s$=stride, $f$ for output channels, and numbers in GRU and FC layers indicate the number of output units. }\medskip
  \label{fig:models}
\end{figure}

We designed two different architectures based on the type of input signal used by the model: AIR-VAD uses only the ACM, and BC-VAD uses only the BCM. From Fig. \ref{fig:models}, we can see that both models have a conventional convolutional-recurrent architecture. The input features are log-Mel-spectrograms, followed by 1D convolutions operating on the frequency axis, and Gated Recurrent Units (GRU) that learn time-dependent information about the input signal. The last two layers of both models are Fully Connected (FC) with ReLU activation for the hidden layer and sigmoid activation to obtain a VAD prediction in the range $[0, 1]$. Due to the inherent eternal noise attenuation of the BCM, the BC-VAD model can have a reduced size ($\approx $ 5k parameters) compared to the AIR-VAD ($\approx$ 58k parameters).

\subsection{VAD baseline for bone conduction}
\label{sec:dsp_vad}

As a baseline for VAD applied on BC signals, we have chosen the method from \cite{DSP-VAD}, which will be called DSP-VAD.

During the reimplementation of DSP-VAD, we identified a potential typo in Eq. (12) of its noise variance updates: DSP-VAD uses a noise estimator to account for additive thermal noise in BC speech. Using the same notation from \cite{DSP-VAD}, the BC signal in the time-frequency domain is:
\begin{equation}
    Y_{BC}(k, l) = S(k, l) + N(k, l)
\end{equation}
where $l$ and $k$ are the time-frame and frequency bin indices, respectively. It is assumed that the speech ($S$) and the thermal noise ($N$) have coefficients with a complex Gaussian distribution. To compute the DSP-VAD's decision on the current frame, the noise variance needs to be estimated ($\hat{\gamma}_N(k)$). In \cite{DSP-VAD} $\hat{\gamma}_N(k)$ is updated when the frame VAD decision rule outputs 1 (speech is detected in the current frame), with the following formula:
\begin{equation}
    \hat{\gamma}_N(k, l) = \alpha_0 \hat{\gamma}_N(k, l - 1) + (1 - \alpha_0)|Y(k, l)|^2
\end{equation}
and remains the same when the VAD does not detect speech. We believe instead that the noise variance should be updated when there is no speech and remain unchanged when there is speech in the current frame. The evaluations made in this work for DSP-VAD are obtained using our revised version of the algorithm from \cite{DSP-VAD}.

\subsection{Evaluation metrics}

We evaluated VAD performance using the Detection Cost Function (DCF ($\downarrow$)): $\textrm{DCF} = 0.75 \cdot \textrm{MR} + 0.25 \cdot \textrm{FAR}$, $\textrm{MR}$ and $\textrm{FAR}$ being miss rate and false alarm rate as defined by the NIST\footnote{\url{https://www.nist.gov/itl/iad/mig/nist-open-speech-activity-detection-evaluation}}. We use two additional metrics adopted from binary classification: the binary accuracy (ACC ($\uparrow$)) with a threshold equal to 0.5 and the area under the receiver operating characteristic curve (AUC ($\uparrow$)), which is threshold-independent. ($\uparrow$) indicates higher is better, and ($\downarrow$) indicates lower is better.

        
        

\section{Experimental setup}


\subsection{Data collection} \label{section:data_acquisition}

All experiments were performed on 16kHz sampled data. 
We collected in-house three datasets with parallel bone conduction and air conduction signals, using two earbud demo kits from Sonion and one reference ACM in front of the participants' mouths. 
Since each demo kit can record using only one modality, the right ear-bud was used to acquire the BC signal and the left ear-bud for AC. The three datasets are:

\textit{Speech subset} contains the target speaker's voice only. We recorded speech from 20 participants (10 males, 10 females) that were reading text in 5 languages (English 8, French 8, Italian 2, Spanish 1, Portuguese 1). In total, we collected approximately 2 hours of speech.
    
\textit{Noise subset} contains external noise only: 10 participants listened to 5 minutes each of noise clips taken from the DNS challenge \cite{DNS_challenge}, without speaking. The noise clips were played using two loudspeakers, placed approximately 1.5 m away from the participant.

\textit{Distractor speech subset} contains external speech only: 10 participants listened to 5 minutes of speech clips taken from the DNS challenge, without speaking and with the same setup as the \textit{noise subset}. The reference air microphone was placed next to the participant's right ear for the last two subsets since we recorded only external sounds.

\subsection{Data preprocessing}
\label{section:data_processing}

Following a procedure similar to \cite{QUT-NOISE-TIMIT}, we randomly cropped and concatenated the speech recordings from the \textit{speech} subset, avoiding cropping consecutive frames in which speech was detected, to form 30 seconds long audio clips. Each clip obtained with this procedure contained speech segments from a single speaker. The  \textit{noise} and \textit{distractor speech} recordings were instead simply concatenated to form clips of the same length (30 seconds).
The obtained dataset is balanced -- evenly divided between clips with low $(<25\%)$, medium (between 25\% and 60\%), and high speech content $(>60\%)$.

After cropping and concatenating the original recordings, the augmented dataset was split as follows: the training subset contained 160 hours of speech from 16 speakers and 160 hours of noise and external speech from 8 participants, and the  test subset included 5 hours of speech from the remaining four speakers and 5 hours of both noise and external speech from 2 participants.


\subsection{Feature extraction}
\label{feature_extraction}

We assume that noisy signals captured by the BCM are composed of the target speaker's voice and noise with external speech.

\begin{equation}
    \label{eq:noisy_bc}
    y_{BC}(t) = s_{BC}(t) + \eta _{BC}(t)
\end{equation}

\begin{equation}
    \eta _{BC}(t) = e_{BC}(t) + \tilde{\eta} _{BC}(t)
\end{equation}
where $y_{BC}(t)$ is the noisy mixture, $s_{BC}(t)$ is the target user's speech, $e_{BC}(t)$ is external speech and $\tilde{\eta} _{BC}(t)$ is external noise (all captured using the BCM).

The clean speech from the target user ($s_{AIR}(t)$), recorded using the ACM, is used to generate the VAD labels:

\begin{equation}
\label{eq:vad_label}
    \mbox{VAD}(n) = \begin{cases} 1, & \mbox{if } ||S_{AIR}(n)|| > T \\ 0, & \mbox{otherwise }  \end{cases}
\end{equation}

\begin{equation}
\label{eq:threshold}
 T = \min _n (||S_{AIR}(n)||) + \alpha \cdot \underset{n}{\mathrm{avg}} (||S_{AIR}(n)||)
\end{equation}

\noindent with $\alpha = 0.3$, the vector $S_{AIR}(n)$ is obtained by STFT of $s_{AIR}(t)$, with 20 ms frame size, $50\%$ overlap and 512 samples FFT. The threshold rule of equation \eqref{eq:threshold} is taken from \cite{sahoo2018voice}. To smooth the target labels, a causal averaging filter is applied to $\mbox{VAD}(n)$ with a length of 0.2 s. $s_{BC}(t)$ and $\tilde{\eta}_{BC}(t)$ are mixed with $SNR \sim \mathcal{N}(15, 5)$ dB in training, and rescaled to signal levels with $\mathcal{N}(-28, 10)$ dBFS. We mapped the frequency range $[50, 2000]$ Hz from the noisy BC mixture into 32 frequency bins in Mel-scale, taking the logarithm of the magnitude of the rescaled spectrogram.


\subsection{Training data for AIR-VAD}
\label{sec:air-vad-training}

Previous sections describe how to obtain the training/test data for BC-VAD. For AIR-VAD, instead, using a procedure similar to Sec.~\ref{section:data_processing}, we have concatenated speech clips from the VCTK dataset \cite{VCTK} to form 30 seconds audio signals. The same operations were performed for noise from the DNS challenge \cite{DNS_challenge}. The training dataset contains 150 hours of speech and 150 hours of noise at a 16 kHz sampling rate. The target VAD labels were computed using Eq.~\eqref{eq:vad_label} on the clean AC speech. 
The augmented input audio signals were obtained by mixing clean speech and noise with $\textrm{SNR} \sim \mathcal{N}(5, 10)$.
The input features were calculated using STFT, with a window size of 20 ms (512 samples after padding the current window with zeros), 50\% overlap. The frequency bins of the resulting spectrogram, ranging from 150 to 5000 Hz, are mapped into 64 Mel-frequency bins. We finally take the logarithm of the magnitude of the scaled spectrogram.

\subsection{Training procedure}
\label{training_procedure}

As VAD is a binary classification task, the Binary Cross Entropy (BCE) loss was used:
\begin{equation}
    \mathcal{L}_{BCE}(z) = - \frac{1}{N} \sum_n z_n \log(\hat{z}_n) + (1 - \hat{z}_n) \log (z_n)
\end{equation}
where $z_n$ and $\hat{z}_n$ are the VAD labels and model's prediction, and $N$ is the number of frames. We used the Adam optimizer \cite{Adam} to update the model's weights with an initial learning rate of 0.001. Each training epoch consisted of 2000 update steps, with a batch size of 8. The learning rate was halved after three consecutive epochs without improvement on the test set, and training was stopped after five consecutive epochs without improvement. This training procedure was applied for both AIR-VAD and BC-VAD.

\section{Results}


We first evaluate the AIR-VAD in Sec.~\ref{sec:air-vad}, and then the BC-VAD in Sec.~\ref{sec:bc-vad} including its complexity analysis in Sec.~\ref{sec:complexity}.

\subsection{VAD for AC}
\label{sec:air-vad}


The proposed AIR-VAD model was evaluated using the QUT-NOISE-TIMIT~\cite{QUT-NOISE-TIMIT} corpus, which contains 600 hours of audio recordings with different noise types and SNR levels. We compared our method with Sohn's Likelihood Ratio Test untrained technique~\cite{Sohn1999}, and three trained neural network-based techniques: Segbroeck~\cite{Segbroeck2013}, Neurogram \cite{WissamA.Jassim2018}) and Spiking Neural Networks (SNN h2 variant) ~\cite{Martinelli2020}. The latter three models were trained on the subset of QUT-NOISE-TIMIT called group A. All models, including AIR-VAD, were evaluated on group B.

Table~\ref{table:air_vad} shows the obtained performance, where we can see that SNN h2 slightly outperforms AIR-VAD in terms of DCF\% in high SNR scenarios, but our proposed method is significantly more robust than the baselines for noisy audio tracks (SNR $\leq $ 5 dB).

\begin{table}[ht!]
\caption{DCF\% ($\downarrow$) scores for different VAD systems relying on air conduction microphone signal on QUT-NOISE-TIMIT}
\vspace{4.5pt}
\label{table:air_vad}
\centering
\begin{tabular}{lccccc}
\toprule
Method / SNR & \textbf{+15} & \textbf{+10} & \textbf{+5}  & \textbf{0}   & \textbf{-5}\\
\midrule
Sohn~\cite{Sohn1999} & 11.1 & 13.4 & 19.7 & 25.9 & 31.3\\
Segbroeck~\cite{Segbroeck2013} & 6.1 & 6.0 & 10.4 & 10.8 & 18.3 \\
Neurogram~\cite{WissamA.Jassim2018} & 5.5 & 5.9 & 10.2 & 10.0 & 17.5\\
SNN h2~\cite{Martinelli2020} & \textbf{3.9} & \textbf{4.5} & 6.2 & 9.4 & 14.1\\
\midrule
AIR-VAD & 5.2 & 5.1 & \textbf{6.1} & \textbf{8.3} & \textbf{10.4}\\
\bottomrule
\end{tabular}
\vspace{-9pt}
\end{table}

\subsection{VAD for BC}
\label{sec:bc-vad}

We evaluated our proposed BC-VAD system in terms of AUC, accuracy (ACC), and DCF\%. We performed these evaluations at different SNR levels for the BCM (${\textrm{SNR}_{\textrm{BC, dB}}}$), using the test dataset from Sec.~\ref{section:data_processing}, which includes non-stationary noise and interference speech. Additionally, we evaluated the performance of AIR-VAD on bone conduction data, simulating a condition when readers use off-the-shelf VAD for BC signals.

\begin{figure}[h]
  \centering
  \centerline{\includegraphics[width=8cm]{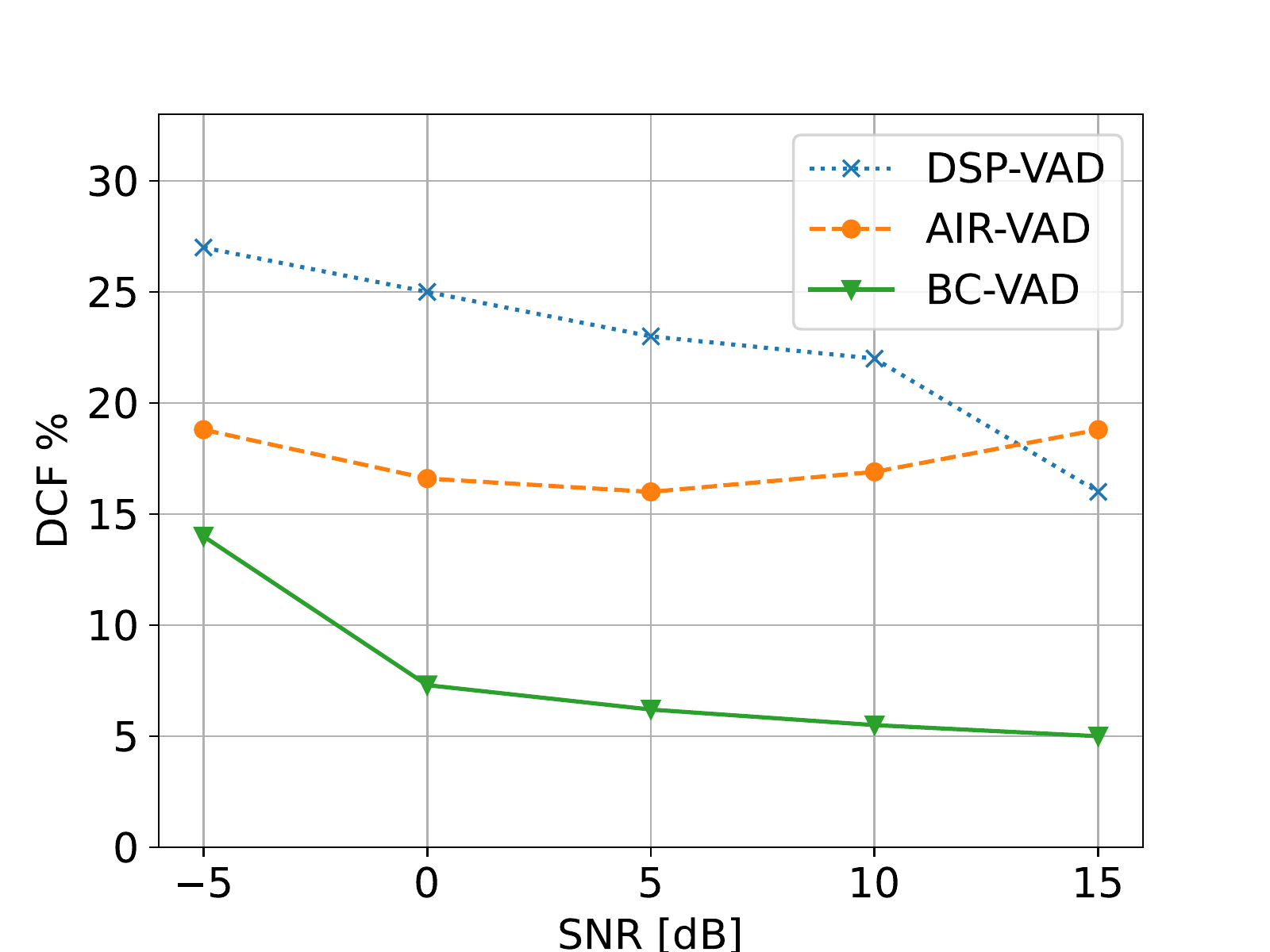}}
  \caption{DCF\% of DSP-VAD, AIR-VAD and BC-VAD on BC speech with non-stationary noise and external speech at various SNR levels}
  \label{fig:dcf-vad}
   \medskip
\end{figure}

Table \ref{table:metrics_bc_vad} shows that BC-VAD performs well in high SNR scenarios ($\geq$ 5 dB), which are the most common in a BCM due to its natural noise attenuation. The baseline DSP-VAD did not achieve good performance due to the various noise types (often non-stationary) and interference speakers of our test dataset, which are more challenging than thermal noise. 
Also, AIR-VAD performed significantly worse than BC-VAD, showing that a state-of-the-art VAD method for AC signals does not necessarily perform well on BC signals. Fig.~\ref{fig:dcf-vad} shows the DCF for the evaluated models, considering various SNR scenarios. Again, BC-VAD performs better than AIR-VAD and DSP-VAD for all SNR levels.

\begin{table}[ht!]
\caption{AUC ($\uparrow$) and ACC ($\uparrow$) results on the BC test dataset containing non stationary noise (Sec.~\ref{section:data_processing}), for different SNR levels of the BCM}
\vspace{4.5pt}
\label{table:metrics_bc_vad}
\centering
\begin{tabular}{lccccc}
\toprule
${SNR_{BC, dB}}$ & \textbf{+15} & \textbf{+10} & \textbf{+5}  & \textbf{0}   & \textbf{-5}\\
\midrule
\textit{\textbf{DSP-VAD:}}\\
AUC & 0.76 & 0.57 & 0.55 & 0.54 & 0.52 \\
ACC & 0.76 & 0.61 & 0.58 & 0.56 & 0.55\\
\midrule
\textit{\textbf{AIR-VAD:}}\\
AUC & 0.93 & 0.93 & 0.93 & 0.91 & 0.85 \\
ACC & 0.84 & 0.85 & 0.85 & 0.83 & 0.76\\
\midrule
\textit{\textbf{BC-VAD:}}\\
AUC & \textbf{0.99} & \textbf{0.98} & \textbf{0.98} & \textbf{0.97} & \textbf{0.95}\\
ACC & \textbf{0.95} & \textbf{0.94} & \textbf{0.94} & \textbf{0.93} & \textbf{0.88}\\
\bottomrule
\end{tabular}
\vspace{-9pt}
\end{table}

\begin{table}[ht!]
\caption{Comparison of DCF\% ($\downarrow$) and Accuracy ($\uparrow$) for the  DSP-based baseline (DSP-VAD) and proposed method (BC-VAD) on clean and noisy signals from the BCM at 20 dB SNR}
\vspace{4.5pt}
\label{table:dsp_vad}
\centering
\begin{tabular}{lcccc}
\toprule
Noise Types & \multicolumn{2}{c}{\textbf{DSP-VAD}} & \multicolumn{2}{c}{\textbf{BC-VAD}} \\
& DCF\% & ACC & DCF\% & ACC\\
\midrule
Clean & 6.90 & 0.88 & \textbf{4.19} & \textbf{0.95} \\
White Noise & 11.5 & 0.88 & \textbf{4.93} & \textbf{0.95}\\
Non-stationary noise & 12.8 & 0.84 & \textbf{4.58} & \textbf{0.95}\\
\bottomrule
\end{tabular}
\vspace{-9pt}
\end{table}

Table~\ref{table:dsp_vad} reports a comparison of BC-DSP and BC-VAD for different noise types. We mixed the speech segments from our BC test subset with three types of noise: white noise at 20 dB SNR, non-stationary noise, distractor speech at 20 dB SNR, and no additional noise (clean speech only). The signal levels were normalized to -28 dBFS. As expected, DSP-VAD performs better in the clean and white noise scenarios, but it is outperformed in all conditions by BC-VAD for accuracy and DCF\%.




\subsection{Complexity analysis}
\label{sec:complexity}

We deployed the BC-VAD on an ARM-Cortex M33 (using the development board NRF5340 Audio DK) to estimate the latency and memory footprint.
Using Tensorflow and Tensorflow-Lite for Microcontrollers, we first quantized the model's weights to 8-bit integers, resulting in a small drop in accuracy (-0.03 in ACC on our BC test dataset at SNR=15 dB) but reducing the computation time and memory requirement of the system. The model occupies a memory region of approximately 35 KB to store the weights, computational graph, inputs, and outputs. With the STFT computation, BC-VAD has an inference time of 2.9 ms for one frame and runs every 10 ms. It thus meets the real-time requirement.

\section{Conclusions}

We have proposed an efficient BC-VAD system that relies on bone conduction signal and is robust against non-stationary noise and interference speech. Our method runs in real-time on an ARM-Cortex M33 processor and achieves better performance than the baseline regarding the detection cost function, the area under the curve, and accuracy. 
We have also shown that a larger version of our model achieves state-of-the-art performance on the QUT-NOISE-TIMIT dataset, relying only on the air-conducted audio signal.



\bibliographystyle{IEEEbib}
\bibliography{myBibliography}

\end{document}